\documentclass[10pt]{article}
\usepackage[mathscr]{eucal}
\usepackage{amssymb,latexsym}
\usepackage{verbatim}
\usepackage{amsmath}
\usepackage{amsthm}
\usepackage{enumerate}
\usepackage{authblk}
\usepackage{color}
\usepackage{url}

\usepackage{tikz}
\usetikzlibrary{matrix}
\usetikzlibrary{snakes}
\usetikzlibrary{arrows,calc,shapes,decorations.pathreplacing}

\usepackage[normalem]{ulem}

\usepackage{amscd,amssymb,amsthm,verbatim}
\usepackage{enumerate}
\usepackage{bm}
\usepackage{mathrsfs, graphicx} 
\usepackage{stmaryrd}

\usepackage[%
bookmarks=true,
colorlinks,
linkcolor=blue,
urlcolor=blue,
citecolor=blue,
plainpages=false,
pdfpagelabels,
final,
breaklinks=true\between
]{hyperref}
\usepackage{hyperref}

\newtheorem{thm}{Theorem}
\newtheorem{lem}[thm]{Lemma}
\newtheorem{Def}[thm]{Definition}
\newtheorem{prop}[thm]{Proposition}
\newtheorem{cor}[thm]{Corollary}

\renewcommand\l{\lambda}

\newcommand\wt{\widetilde}

\renewcommand\S{\Sigma}

\newcommand\e{\epsilon}
\renewcommand\b{\beta}

\renewcommand\div{{\rm div}}

\renewcommand\l{\lambda}
\newcommand\g{\gamma}

\renewcommand\a{\alpha}

\newcommand\beq{\begin{equation}}
\newcommand\eeq{\end{equation}}
\newcommand\ben{\begin{enumerate}}
\newcommand\een{\end{enumerate}}
\newcommand\bit{\begin{itemize}}
\newcommand\eit{\end{itemize}}

\DeclareMathOperator{\diver}{div}

\renewcommand{\div}{\diver}

\newcommand{\red}[1]{\textcolor{red}{#1}}

\newcommand{\R}{\mathbb R}

\newcommand{\ov}{\overline}

\newcommand{\pd}{\partial}

\newcommand{\Z}{\mathbb{Z}}

\newcounter{mnotecount}
\renewcommand{\themnotecount}{\thesection.\arabic{mnotecount}}

\newcommand{\mnote}[1]
{\protect{\stepcounter{mnotecount}}$^{\mbox{\footnotesize $
      \bullet$\themnotecount}}$ \marginpar{
\raggedright\tiny\em
    $\!\!\!\!\!\!\,\bullet$\themnotecount: #1} }

\title{Topological censorship in spacetimes compatible with $\Lambda > 0$}

\author[1]{Martin Lesourd\footnote{mlesourd@math.harvard.edu}}
\author[2]{Eric Ling\footnote{eling@math.rutgers.edu}}

\affil[1]{Black Hole Initiative, Harvard University, Cambridge, MA}

\affil[2]{Rutgers University, New Brunswick, NJ}

\begin{document}
\date{}
\maketitle

\vspace{.15in}

\begin{abstract} 
Currently available topological censorship theorems are meant for gravitationally isolated black hole spacetimes with cosmological constant $\Lambda=0$ or $\Lambda<0$. Here, we prove a topological censorship theorem that is compatible with $\Lambda>0$ and which can be applied to whole universes containing possibly multiple collections of black holes. The main assumption in the theorem is that distinct black hole collections eventually become isolated from one another at late times, and the conclusion is that the regions near the various black hole collections have trivial fundamental group, in spite of there possibly being nontrivial topology in the universe.
\end{abstract}

\newpage

\section{Introduction}
Topological censorship theorems show, under various physically motivated assumptions, that the topology of the region exterior to a black hole (Domain of Outer Communication, `DOC') is restricted in various respects. The DOC is defined in terms of the conformal boundary, $\mathcal{J}$, which is associated to the spacetime. The causal character of $\mathcal{J}$ is taken to be timelike, null, or spacelike depending on whether these spacetimes model solutions to the Einstein field equations 
\[
\textbf{G}\equiv \textbf{Ric}-\frac{1}{2}\textbf{g} \: R=\textbf{T}-\Lambda\: \textbf{g}
\]
with cosmological constant $\Lambda<0,=0,>0$, respectively. So far, topological censorship theorems have been proven in the context of $\Lambda=0$ and $\Lambda<0$. In view of the currently preferred $\Lambda$CDM model which supports a cosmological constant $\Lambda > 0$ \cite{Planck}, we shall seek to extend these theorems to this setting.

Typical topological censorship theorems are based on the following assumptions:
\begin{itemize}
\item[-] the null energy condition,
i.e. $\textbf{Ric}(\textbf{n},\textbf{n})\geq 0$ for all null vectors $\textbf{n}$ (various averaged versions would also suffice),
\item[-] global hyperbolicity of the DOC,
\item[-] a topological assumption concerning $\mathcal{J}$.
\end{itemize} 
\hspace{0.1in} In the asymptotically flat setting ($\Lambda = 0$), the conformal boundary is a disjoint union, $\mathcal{J}=\mathcal{J}^+ \cup \mathcal{J}^-$, of null hypersurfaces $\mathcal{J}^+$ and $\mathcal{J}^-$. Both of these are topologically $\mathbb{R}\times S^2$, the DOC is defined as $\text{DOC} = I^+(\mathcal{J}^-)\cap I^-(\mathcal{J}^+)$, and the main theorem in this setting is that the DOC is simply connected \cite{CG19, FSW93, G95, G96, GW96}.
In the more complicated case of timelike conformal boundaries, one obtains topological restrictions on the DOC involving the genus $g$ of the surface `at infinity', and the genus $g_i$ of the horizon components \cite{GSWW99}. See also \cite{CGS} and \cite{GGL}.\\ \indent 
Here, motivated by black hole solutions with $\Lambda>0$ (eg.\@ Kerr-de Sitter) and today's preferred cosmological models, we prove a topological censorship theorem for spacetimes compatible with $\Lambda>0$ with the following features.

\medskip

\begin{itemize}
\item{Although the theorem is intended to cover spacetimes that admit a future (and or past) spacelike conformal boundary $\mathcal{J}$, the theorem is not actually stated in terms of $\mathcal{J}$. The event horizon (which is normally defined by the boundary of the timelike past of $\mathcal{J}$) is replaced with a suitable Cauchy horizon.}
\item{The theorem applies to spacetimes containing multiple black hole collections, rather than an isolated system like an asymptotically flat spacetime.}
\item{The geometric set up is that the black holes eventually become isolated from one another, and by this it is meant that the intersection between the event horizon and a suitable spacelike hypersurface has multiple connected components, cf.\@ \cite{CMa} for such set-ups. One can think of this assumption as a consequence of certain PDE conjectures like Weak Cosmic Censorship and Final State, which say that generic gravitational collapse leads to a number of distinct black holes regions that eventually become isolated from one another.}
\item{The main assumption of our theorem is that at \emph{late times}, the black holes become so isolated that light from the region near one black hole does not reach the event horizon of another black hole. This assumption is supported by the presence of a positive cosmological constant as shown in Figure \ref{SdS fig}.}
\item{The conclusion is that a local form of topological censorship holds. Namely, the regions in the vicinity of the black holes have trivial fundamental group but the region between the black holes may contain nontrivial topology. In essence this shows that the topology of the universe may be nontrivial but the topology near black holes is trivial.}
\end{itemize} 

\medskip

\subsection{Examples}\label{examples section}
%

\medskip

The following examples show what one can and cannot prove with regards to topological censorship in the asymptotically de Sitter setting. 

\medskip
\medskip

\noindent \textbf{Example 1: de Sitter $\R P^3$ quotient.} This example shows that in the de Sitter setting, it is possible to have a simply connected $\mathcal{J}$ and yet the DOC is \emph{not} simply connected.

 Consider the de Sitter spacetime $(M,g)$ where $M = I \times S^3$ and $g = \cos^{-2}(t)(-dt^2 + d\omega^2)$; here $I = (-\frac{\pi}{2}, \frac{\pi}{2})$ and $d\omega^2$ is the usual round metric on $S^3$. The conformal spacelike boundaries correspond to $\mathcal{J}^+ = \{t = + \frac{\pi}{2}\}$ and $\mathcal{J}^- = \{t = - \frac{\pi}{2}\}$; each are topologically $S^3$.  By quotienting out each $S^3$ by the antipodal identification map, one obtains real projective space $\R P^3 = S^3/ \sim$ and applying this identification to each $\{t\} \times S^3$ for each $t \in \R$ yields a new spacetime $(M_1, g_1)$ where $M_1 = I \times \R P^3$ and $g_1$ is just the metric induced by $g$.  This spacetime inherits the conformal structure from the de Sitter spacetime and so its conformal spacelike boundaries are given by $\mathcal{J}^{\pm}_1 = \mathcal{J}^\pm/\sim$.

\begin{figure}[t]
\begin{center}
\mbox{
\includegraphics[width=3.8in]{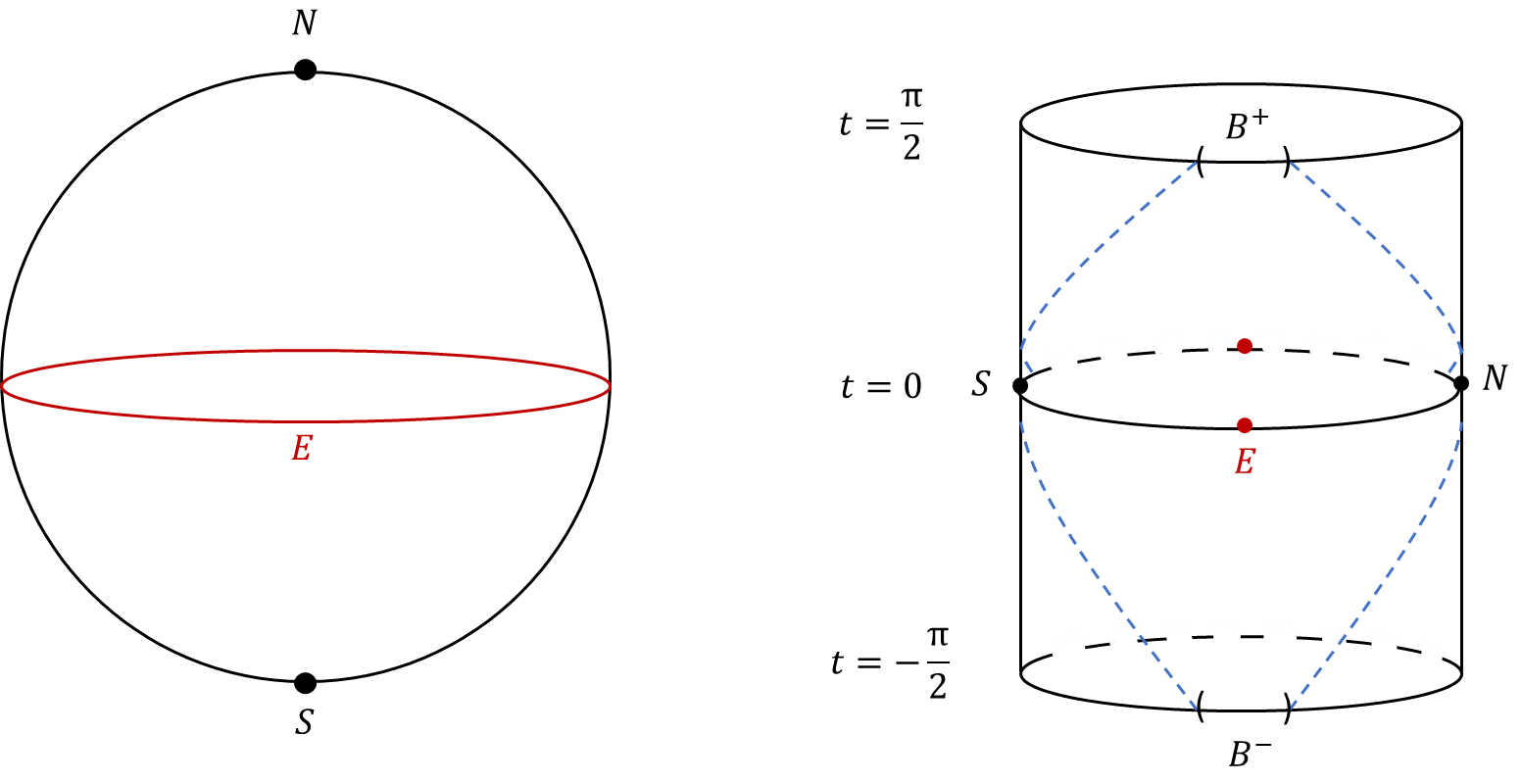}
}
\end{center}
\caption{\small{Left: The sphere $S^3$; the equator $E$ is topologically $S^2$. Identifying antipdoal points on $S^3$ yields $\R P^3$; the equator under this identification becomes an $\R P^2$. Right: The de Sitter $\R P^3$ quotient.  $B^\pm$ are two small open balls in $\mathcal{J}_1^\pm$. The spacetime $M_2 = I^+(B^-) \cap I^-(B^+)$ is not simply connected and yet $\mathcal{J}_2^\pm = B^\pm$ are simply connected.}}
\label{RP3ds fig}
\end{figure}

Now consider the spacetime $(M_2, g_2)$ where $M_2 = I^+(B^-) \cap I^-(B^+)$ where $B^\pm$ are two (small) open balls embedded in $\mathcal{J}^\pm_1$ centered around the same point. See Figure \ref{RP3ds fig}. The spacelike conformal boundaries for $(M_2, g_2)$ are given by $\mathcal{J}_2^\pm = B^\pm$. The spacetime $(M_2, g_2)$ is globally hyperbolic, satisfies the null energy condition, has simply connected spacelike conformal boundaries, and yet the DOC (which equals $M_2$) is not simply connected since its Cauchy surfaces are topologically $\R P^3$. 
  
\medskip
\medskip

\noindent\textbf{Example 2: Schwarzschild-de Sitter spacetime.} Our late time assumption that we consider in our main theorem is motivated by the late time asymptotics in the Schwarzschild-de Sitter spacetime.

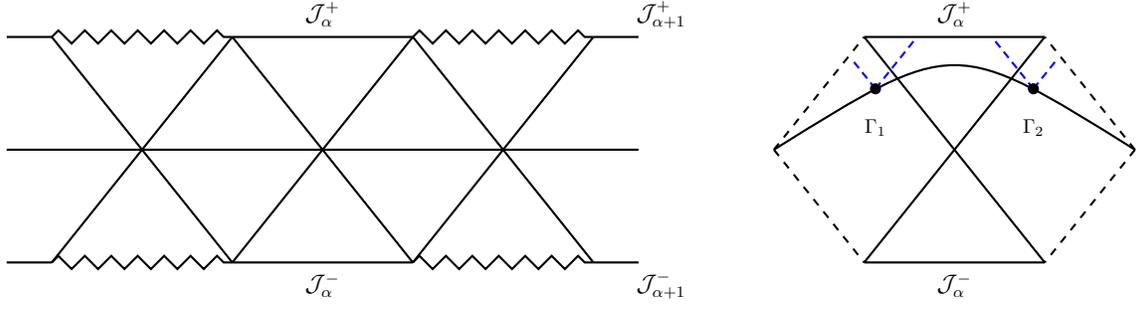
\begin{figure}[t]
\begin{center}
\begin{tikzpicture}[scale = 0.60]


\draw[-,thick] (-5,5) -- (-4,5);
\draw[snake=zigzag,thick] (-4,5) -- (0,5);
\draw[-,thick](0,5) -- (4,5);
\draw[snake=zigzag,thick] (4,5)-- (8,5);
\draw[-,thick] (8,5) -- (9,5);

\draw[-,thick] (-5,0) -- (-4,0);
\draw[snake=zigzag,thick] (-4,0) -- (0,0);
\draw[-,thick](0,0) -- (4,0);
\draw[snake=zigzag,thick] (4,0)-- (8,0);
\draw[-,thick] (8,0) -- (9,0);

\draw[-,thick] (-4,5) -- (0,0);
\draw[-,thick] (-4,0) -- (0,5);
\draw[-,thick] (0,5) -- (4,0);
\draw[-,thick] (0,0) --(4,5);
\draw[-,thick] (4,5) -- (8,0);
\draw[-,thick] (4,0) -- (8,5);

\node(x1) [scale=.75] at (2,5.5) {\large $\mathcal{J}^+_\alpha$}; 
\node(x2) [scale=.75]  at (2,-0.5) {\large $\mathcal{J}^-_\alpha$};
\node(x3) [scale=.75] at (9.5,5.5) {\large $ \mathcal{J}^+_{\alpha+1}$};
\node(x4) [scale=.75] at (9.5,-0.5) {\large $ \mathcal{J}^-_{\alpha+1}$};

\draw[-, thick] (-5,2.5) to (9,2.5);


\draw[-,thick](14,5) -- (18,5);

\draw[-,thick](14,0) -- (18,0);

\draw[-,thick, dashed] (12,2.5) -- (14,0);
\draw[-,thick, dashed] (12,2.5) -- (14,5);
\draw[-,thick] (14,5) -- (18,0);
\draw[-,thick] (14,0) --(18,5);
\draw[-,thick, dashed] (20,2.5) -- (18,5);
\draw[-,thick, dashed]  (20,2.5) -- (18,0);

\node(x5) [scale=.75] at (16,5.5) {\large $\mathcal{J}^+_\alpha$}; 
\node(x6) [scale=.75]  at (16,-0.5) {\large $\mathcal{J}^-_\alpha$};

\draw[-, thick] (12,2.5) .. controls (16,5) .. (20,2.5);

\draw [thick, densely dashed, blue] (14.25,3.85) -- (15.17,5);
\draw [thick, densely dashed, blue] (14.25,3.85) -- (13.75, -1.25*13.75 + 21.6625);
\node [scale = .4] [circle, draw, fill = black] at (14.25, 3.85)  {};
\draw (14.25,3) node [scale =.75] {$\Gamma_1$};

\draw [thick, densely dashed, blue] (17.75,3.85) -- (16.83,5);
\draw [thick, densely dashed, blue] (17.75,3.85) -- (18.25, 1.25*18.25 + -18.3375);
\node [scale = .4] [circle, draw, fill = black] at (17.75, 3.85)  {};
\draw (17.75,3) node [scale =.75] {$\Gamma_2$};

\end{tikzpicture}
\end{center}
\caption{\small{Left: The maximal analytic extension $(M_*, g_*)$ of the Schwarzschild-de Sitter spacetime. The horizontal line in the middle represents a Cauchy surface which has topology $\R \times S^2$. Right: The $\text{DOC}_\a$. A particular late time Cauchy surface is shown. Note that the two spheres $\Gamma_1$ and $\Gamma_2$ do not have intersecting future lightcones. This will play a role in our late time assumption in our theorem.}}
\label{SdS fig}
\end{figure}

Fix $m > 0$ and $\Lambda > 0$ such that $9\Lambda m^2 < 1$. Let $f(r) = 1 - \frac{2m}{r} -\frac{\Lambda}{3}r^2$. Then $f(r)$ has two positive roots $r_1 < r_2$.  Let $(M,g)$ denote the spacetime 
\[
M \,=\, \R \times (r_1, r_2) \times S^2 \:\:\:\: \text{ and } \:\:\:\: g \,=\, -f(r)dt^2 + \frac{1}{f(r)}dr^2 + r^2d\Omega^2
\]
where $d\Omega^2$ is the usual round metric on $S^2$. The roots $r_1$ and $r_2$ correspond to coordinate singularities known as the \emph{event horizon} and \emph{cosmological horizon}, respectively. Let $(M_*, g_*)$ denote the maximal analytic extension of $(M,g)$. A Penrose diagram for $(M_*, g_*)$ is given in Figure \ref{SdS fig}. For a construction of the maximal analytic extension, see section 6.3 in \cite{ChruBook}. Note that $(M_*, g_*)$ is globally hyperbolic with Cauchy surfaces topologically $\R \times S^2$.  In $(M_*,g_*)$ there are countably infinite black hole regions which are separated by countably infinite connected components of the spacelike conformal boundary $\mathcal{J}^{\pm} = \bigsqcup_{\a \in \Z} \mathcal{J}^\pm_\a$ which satisfy $\mathcal{J}^-_\a \subset I^-(\mathcal{J}^+_\a)$ and $\mathcal{J}^-_\a$ is disjoint from $I^-(\mathcal{J}^+_\b)$ for all $\b \neq \a$. Moreover $\mathcal{J}^\pm_\a$ is topologically $\R \times S^2$ for each $\a$. 

Let's focus on a specific $\a$ and consider the $\text{DOC}_\a = I^+(\mathcal{J}^-_\a) \cap I^-(\mathcal{J}^+_\a)$. The Cauchy surfaces for $\text{DOC}_\a$ are also topologically $\R \times S^2$. A particular \emph{late time} Cauchy surface is shown in Figure \ref{SdS fig}. The spheres $\Gamma_1$ and $\Gamma_2$ can be thought of spheres surrounding two distinct black holes 
in the spacetime. Note that for this particular late time Cauchy surface, the future lightcones of $\Gamma_1$ and $\Gamma_2$ do not intersect. This property - that the future lightcones of $\Gamma_1$ and $\Gamma_2$ do not intersect - motivates our ``late time assumption" which appears in our theorem.

%
%
%

%
%
%

\medskip
\medskip

\noindent \textbf{Example 3: Schwarzschild-de Sitter geon.}
Our theorem proves that the regions near black holes are topologically trivial while still allowing for nontrivial topology in the universe as a whole. This example demonstrates one scenario where the topology around a black hole is trivial even though the universe as a whole contains nontrivial topology.

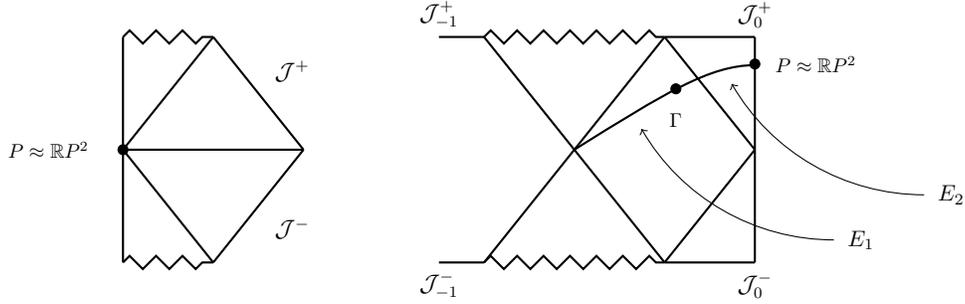
\begin{figure}[t]
\begin{center}
\begin{tikzpicture}[scale = 0.60]


\draw[snake=zigzag,thick] (-12,5) -- (-10,5);

\draw[snake=zigzag,thick] (-12,0) -- (-10,0);

\draw[-,thick] (-12,2.5) -- (-10,0);
\draw[-,thick] (-12,2.5) -- (-10,5);
\draw[-,thick] (-10,5) -- (-8,2.5);
\draw[-,thick] (-10,0) --(-8,2.5);

\draw[-,thick] (-12,0) -- (-12,5);

\draw[-,thick] (-12,2.5) -- (-8,2.5);

\node [scale=.75] at (-8.25,4.25) {\large $\mathcal{J}^+$}; 
\node [scale=.75] at (-8.25,0.75) {\large $\mathcal{J}^-$}; 

\node [scale = .4] [circle, draw, fill = black] at (-12, 2.5)  {};
\draw (-13.65, 2.5) node [scale =.75] {$P \approx \R P^2$};


\draw[-,thick] (-5,5) -- (-4,5);
\draw[snake=zigzag,thick] (-4,5) -- (0,5);
\draw[-,thick](0,5) -- (2,5);

\draw[-,thick] (-5,0) -- (-4,0);
\draw[snake=zigzag,thick] (-4,0) -- (0,0);
\draw[-,thick](0,0) -- (2,0);

\draw[-,thick] (-4,5) -- (0,0);
\draw[-,thick] (-4,0) -- (0,5);
\draw[-,thick] (0,5) -- (2,2.5);
\draw[-,thick] (0,0) --(2,2.5);

\draw[-, thick] (-2,2.5) .. controls (2,5) .. (6,2.5);
\fill[white] (2,0) -- (6,0) -- (6,5) -- (2, 5) -- (2,0);

\draw [->] (3.75,.5) arc [start angle=-90, end angle=-150, radius=140pt];
\draw (4.35,.5) node [scale =.85] {$E_1$};

\draw [->] (5.75,1.5) arc [start angle=-90, end angle=-150, radius=140pt];
\draw (6.35,1.5) node [scale =.85] {$E_2$};

\draw[-,thick] (2,0) -- (2,5);

\node(x1) [scale=.75] at (2,5.5) {\large $\mathcal{J}^+_0$}; 
\node(x2) [scale=.75]  at (2,-0.5) {\large $\mathcal{J}^-_0$};
\node(x3) [scale=.75] at (-5,5.5) {\large $\mathcal{J}^+_{-1}$}; 
\node(x4) [scale=.75]  at (-5,-0.5) {\large $\mathcal{J}^-_{-1}$};

\node [scale = .4] [circle, draw, fill = black] at (0.25, 3.85)  {};
\draw (0.25,3.15) node [scale =.75] {$\Gamma$};

\node [scale = .4] [circle, draw, fill = black] at (2, 4.4)  {};
\draw (3.35,4.4) node [scale =.75] {$P \approx \R P^2$};

\end{tikzpicture}
\end{center}
\caption{\small{Left: The Schwarzschild $\R P^3$ geon. The horizontal line in the middle is a Cauchy surface with nontrivial topology due to the surface $P$. Right: The Schwarzschild-de Sitter $\R P^3$ geon. The Cauchy surface $E = E_1 \cup E_2$ (which is separated by $\Gamma = E_1 \cap E_2$) for $\text{DOC}_0$ is topologically nontrivial. The region $E_1$ near the black hole \emph{is} topologically trivial; the nontrivial topology in $E$ occurs at the surface $P$ in $E_2$ which is far away from the black hole. }}
\label{SdS geon fig}
\end{figure}

 The Schwarzschild $\mathbb{R}P^3$ geon described in \cite{FSW93} arises from the following construction: Fix $m > 0$. Let $(M_*, g_*)$ denote the maximal analytic extension of the Schwarzschild spacetime. In Kruskal coordinates, we have
\[
M_* \,=\, D \times S^2 \:\:\:\: \text{ and } \:\:\:\: g_* \,=\, \frac{32 m^3}{r}e^{-r/2m}(-dT^2 + dX^2) + r^2d\Omega^2 
\]
where $D = \{(T,X) \mid X \in \R \text{ and } T^2 - X^2 < 1 \}$ and $r$ is a function of $X^2 - T^2$. The \emph{Schwarzschild $\R P^3$ geon} is the spacetime $(M,g)$ obtained by making the following identifications: $X \sim -X$ and $p \sim -p$ for $p \in S^2$. One can imagine that one is ``folding over" $(M_*, g_*)$ to yield $(M,g)$. In Figure \ref{SdS geon fig}, we fold from left to right. We note that $(M,g)$ has only one asymptotic end (as oppose to two in $(M_*, g_*)$), the minimal horizon (denoted by $P$ in Figure \ref{SdS geon fig}) in the $t = 0$ slice is topologically $\R P^2$ (as oppose to $S^2$ in $(M_*, g_*)$), and the $t = 0$ slice is topologically $\R P^3 \setminus \{\text{pt}\}$ (as oppose to $\R \times S^2$ in $(M_*, g_*)$).  

Now let $(M_*, g_*)$ denote the maximal analytic extension of the Schwarzschild-de Sitter spacetime. This spacetime has a similar symmetry as in the maximal analytic extension of the Schwarzschild spacetime. But this time we ``fold over"  at the intersection of the cosmological horizons as oppose to the event horizons. The resulting spacetime $(M,g)$ is the \emph{Schwarzschild-de Sitter $\R P^3$ geon} which is illustrated in Figure \ref{SdS geon fig} where we have folded from right to left. A Cauchy surface $E = E_1\cup E_2$ for $\text{DOC}_0$ is shown in Figure \ref{SdS geon fig}. Note that $\Gamma$ separates $E$ so that $\Gamma = E_1 \cap E_2$. Analogous to the $t = 0$ slice in the Schwarzschild $\R P^3$ geon, $E$ is topologically $\R P^3 \setminus\{ \text{pt}\}$. But notice that $E_1$ is topologically trivial, while $E_2$ is topologically nontrivial since it contains the surface $P$ which is topologically $\R P^2$. This shows that the region near the black hole (e.g. $E_1$) is topologically trivial even though when can have nontrivial topology in the universe far away from the black hole (e.g. $E_2$).

Lastly, in Figure \ref{SdS geon fig}, we have $\mathcal{J}_0^\pm \approx \R P^3 \setminus \{\text{pt}\}$ while $\mathcal{J}^\pm_{-n} \approx \R \times S^2$ for all $n = 1, 2, \dotsc$. So we see that the nontrivial topology of $\text{DOC}_0$ coincides with the nontrivial topology of $\mathcal{J}_0^\pm$. However, by removing the $\R P^2$ equator from $\mathcal{J}_0^\pm$ and letting $\text{DOC}_0'$ denote the corresponding domain of outer communication, we see that $\text{DOC}_0' = \text{DOC}_0$. But the spacelike conformal boundaries for $\text{DOC}_0'$ are topologically $\R \times S^2$ and hence simply connected. This is analogous to example 1 since again we have a topologically nontrivial domain of outer communications but a topologically trivial spacelike conformal boundary.

\medskip

\subsection{The Theorem} \label{the thm section}

\medskip
Let $(M,g)$ be a spacetime. For our theorem to apply to multiple black holes, we make the following three assumptions on $M$. For the sake of simplicity, we assume all objects are smooth. Figure \ref{set-up fig} shows a picture of the general set up we have in mind.

\medskip 
\begin{itemize} 
\item[(1)] There is a spacelike Cauchy surface $V$ for $M$.

\item[(2)] Let $\mathcal{I}$ be an indexed set. For each $i \in \mathcal{I}$, there is an embedded surface $\Sigma_i$ which separates $V$. Let $B_i'$ and $E_i'$ form a separation for $V \setminus \S_i$. Set $B_i = B'_i \sqcup \S$ and $E_i = E_i' \sqcup \S_i$. Hence
\[
    V \,=\, B_i \cup E_i \:\:\:\: \text{ and } \:\:\:\: \Sigma_i \,=\, B_i \cap E_i. 
\]
We assume $\S_i$ is a closed set so that $B_i$ and $E_i$ are closed sets. We assume each $B_i$ is  diffeomorphic to $\Sigma_i \times [0,\epsilon)$ and represents a collars worth of the black hole region; $\S_i$ represents the boundary of the black hole region. We make no assumptions on the connectedness nor compactness of $\S_i$.

\item[(3)] For each $i \in \mathcal{I}$, there is a smooth embedded 2-sphere $\Gamma_i \subset E_i \setminus \S_i$ which separates $E_i \setminus \S_i$. Let $E_{1,i}'$ and $E_{2,i}'$ form a separation for $E_i \setminus (\S_i \sqcup \Gamma_i)$. Set $E_{1,i} = E_{1,i}' \sqcup \Gamma_i \sqcup \Sigma_i$ and $E_{2,i} = E_{2,i}' \sqcup \Gamma_i$.  Hence 
\[ 
E_i \,=\, E_{1,i} \cup E_{2,i} \:\:\:\: \textnormal{ and } \:\:\:\: \Gamma_i \,=\, E_{1,i} \cap E_{2,i}.
\]
We assume $E_{1,i}$ is connected. We call $E_{1,i}$ and $E_{2,i}$ as the \emph{inward} and \emph{outward} directions of $\Gamma_i$, respectively. We assume each $\Gamma_i$ is inner trapped with respect to $V$. Lastly, the \emph{cosmological core} is defined by $C = \bigcap_i E_{2, i}$. 
\end{itemize}

\medskip
\medskip

\noindent\emph{Remark.} One should think of $\Gamma_i$ as ``enveloping" the black hole region $B_i$. For example, one can imagine that $\Gamma_i$ is a large 2-sphere surrounding one specific black hole, or alternatively, one can imagine that $\Gamma_i$ surrounds an entire galaxy. It should be noted that $E_{2,i}$ contains all the other black hole regions $B_j$ for $j \neq i$. The definition of the cosmological core is motivated by recognizing that $C = \bigcap_i E_{2,i}$ is what is left over after removing the regions $E_{1,i}$ from $V$.

\begin{figure}[t]
\begin{center}
\begin{tikzpicture}[scale = 0.65]

\draw[snake=zigzag, thick] (-5,5) -- (-3,5);
\draw[-, thick] (-3,5) -- (3,5);
\draw[snake=zigzag, thick] (3,5) -- (5,5);

\draw[-, thick] (-3,5) -- (-27/5,2);
\draw[-, thick] (-3,5) -- (-3/5, 2);
\draw[-, thick](3,5) -- (3/5,2);
\draw[-, thick] (3,5) -- (27/5,2);

\draw[-, thick] (-5.25,3) .. controls (0,4.5) .. (5.25,3);

\node [scale = .4] [circle, draw, fill = black] at (-4.4,3.25)  {};
\node(x1) at (-4.15,2.65) [scale = .75] {\large $\Sigma_i$};

\node [scale = .4] [circle, draw, fill = black] at (-2.5,3.75)  {};
\node(x2) at (-2.25,3.15) [scale = .75] {\large $\Gamma_i$};

\draw [->] (-6.25,4.1) arc [start angle=90, end angle=25, radius=40pt];
\node at (-6.75,4.125) [scale = .75] {\large $B_i$};

\draw [->] (-3.1,6.2) arc [start angle=135, end angle=205, radius=65pt];
\node at (-2.5,6.5) [scale = .75] {\large $E_{1,i}$};

\draw [->] (0.5,6.425) arc [start angle=35, end angle=-7.55, radius=90pt];
\node(x3) at (0.075,6.85) [scale = .75] {\large $E_{2,i}$};

\node(x6) at (-0.2,0) [scale = .75] {\large $\text{Cauchy surface} \,=\, V\,=\,B_i \cup E_i\,=\,  B_i \cup \left( E_{1,i} \cup E_{2,i}\right) $};

\end{tikzpicture}
\end{center}
\caption{\small{The set-up for our main theorem. Note that $E_{2,i}$ contains all the other black hole regions $B_{j}$ for $j \neq i$. The conclusion of our theorem is that the regions $E_{1,i}$ near the black hole are simply connected. }}
\label{set-up fig}
\end{figure}

\begin{figure}[t]
\begin{center}
\includegraphics[width=10cm]{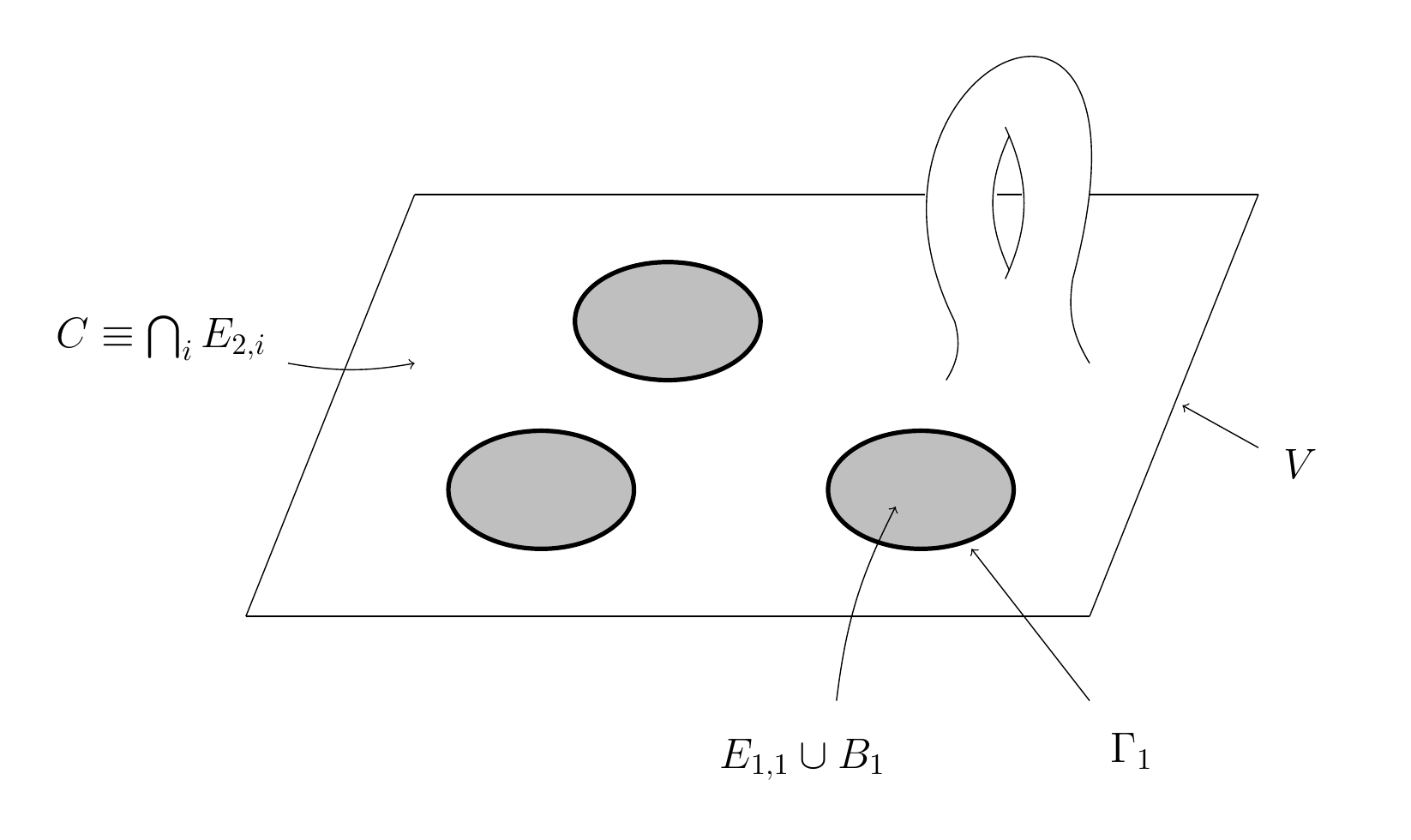}
\end{center}
\caption{\small{Although the regions $E_{1,i}$ near the black hole are simply connected, there can still exist nontrivial topology in the cosmological core $C$ as indicated in this figure above. An example of this occurs in the Schwarzschild-de Sitter geon illustrated in Figure \ref{SdS geon fig}. }}
\label{cos core fig}
\end{figure}

\medskip
\medskip

The late time assumption in our theorem is provided in the following definition.

\medskip

\begin{Def}\label{late time def}
\emph{
Let $(M,g)$ be a spacetime. Let $H^+(S)$ denote the future Cauchy horizon of a closed an achronal set $S$. We say $(M,g)$ has \emph{settled down at late time} if it satisfies assumptions (1) - (3) above and that for all $i$, each inward normal future directed null geodesic $\g$ emanating from $\Gamma_i$ is either future complete or crosses $H^+(E_i)$ (i.e. intersects it transversely). 
}
\end{Def}

\medskip
\medskip

\newpage

\noindent\emph{Remarks.} 
\begin{itemize}
\item[-] Imagine two black holes $\S_1$ and $\S_2$ orbiting around each other on a collision course. Let $\Gamma_1$ and $\Gamma_2$ denote two-spheres surrounding $\S_1$ and $\S_2$. In this case, one can believe that an inward normal future directed null geodesic from $\Gamma_1$ intersects $H^+(E_2)$ and becomes future incomplete without ever crossing $H^+(E_1)$. Hence Definition \ref{late time def} would not hold in this case. This is why we include ``at late time" in our definition. 

\item[-] Spacetimes that are asymptotically de Sitter are expected to settle down at late time since the cosmological constant does not allow for communication between black holes far into the future. This is depicted in the picture on the right in Figure \ref{SdS fig}. In this case it is clear that any null geodesic from $\Gamma_1$ will never meet the event horizon on the right. 

\item[-] Our late time assumption can probably be relaxed in various ways. Perhaps a more reasonable assumption than the one given in Definition \ref{late time def} is that each inward normal future null geodesic $\g$ emanating from $\Gamma_i$ is either future complete or crosses $H^+(E_i)$ or meets a timelike cylinder formed by the integral curves of a timelike vector field starting on $\Gamma_i$. This would leave open the possibility that $\g$ does not cross $H^+(E_i)$ and is incomplete in the future perhaps by crossing another event horizon $H^+(E_j)$ for $j \neq i$, but in order for $\gamma$ to get to $H^+(E_j)$, it must have met the timelike cylinder formed from $\Gamma_i$. This setting would be more in the spirit of \cite{CGS} (see also Section 3.3.4 in \cite{ChruBook}). 

\end{itemize} 

\medskip
\medskip

\noindent The main theorem of our paper is:

\medskip

\begin{thm}\label{main}
Assume $(M,g)$ is a spacetime which has settled down at late time and satisfies the null energy condition. Let $N = |\mathcal{I}|$ (i.e. $N$ is the cardinality of the black hole regions). 
\begin{itemize}
\item[\emph{(a)}] Suppose $N \geq 2$. Then, for each $i$, $\pi_1(E_{1,i})$ is trivial and $\S_i$ is a finite disjoint union of $S^2$s. Moreover $\pi_1(V) = \pi_1(C)$. 
\item[\emph{(b)}] Suppose $N = 1$. Consequently, we drop the subscript $i$. In this case the following hold. 
\begin{itemize}
\item[\emph{(i)}] $\pi_1(E_1)$ is finite.
\item[\emph{(ii)}] If $E_2$ is noncompact, then $\pi_1(E_1)$ is trivial and $\S$ is a finite disjoint union of $S^2$s.
\item[\emph{(iii)}] If $\pi_1(E_2)$ is nontrivial, then $\pi_1(E_1)$ is trivial and $\S$ is a finite disjoint union of $S^2$s.
\item[\emph{(iv)}] If $\pi_1(E_1)$ is nontrivial, then $E_2$ is diffeomorphic to the $3$-disc.
\end{itemize}  
\end{itemize}
\end{thm}

\medskip
\medskip

\noindent\emph{Remarks.}

\begin{itemize}

\item[-] The proof of Theorem \ref{main} is given in section \ref{proof section}. It combines the compactness argument of Lemma \ref{compactness lemma} applied to suitable covering spacetime. Both the compactness lemma and the covering construction are inspired from \cite{BG}, but the details of the set-up and proof differ.

\item[-] The case $N \geq 2$ is more physically reasonable since we observe multiple black holes from all galaxies throughout the universe. Note that in this case we get the desired conclusion that the regions $E_{1,i}$ near each black hole region are topologically trivial. Moreover, we find that the fundamental group of the spacetime is determined by the fundamental group of the cosmological core: $\pi_1(V) = \pi_1(C)$. An example of a situation like this was given by the Schwarzschild-de Sitter $\R P^3$ geon in example 3 of the previous section. See Figure \ref{SdS geon fig}. However, one should recognize that this example really belongs to the case $N = 1$. 

\item[-] Physically, the case $N = 1$ represents the scenario when we can envelop all the black holes in the universe with exactly one (very large) 2-sphere $\Gamma$. Mathematically, it also applies to spacetimes with just one black hole as in Figure \ref{SdS geon fig}.  We do not obtain as many nice conclusions in this case. We note though that if $E_2$ is noncompact, then $\pi_1(E_1)$ is trivial. Case (iv) is interesting since we do not know of any spacetime examples which satisfy the hypothesis of case (iv). Determining whether or not such examples exist is an open question.

\end{itemize}

\vspace{0.2in} \textbf{Acknowledgements.} Martin Lesourd thanks the John Templeton and Gordon Betty Moore foundations for their support of the Black Hole Initiative. Eric Ling thanks the Harold H. Martin Postdoctoral Fellowship. Finally, both authors would like to express their thanks to Greg Galloway, with whom we discussed examples which greatly improved our understanding.

\medskip
\medskip

\section{Compactness Lemma}\label{compactnesssection}

\medskip

\subsection{Causal theory preliminaries}

We state some results from causal theory used in our set-up. Standard references for these results are \cite{ChruBook,  HE, ON, Wald}. 

A \emph{spacetime} is a pair $(M,g)$ where $M$ is a smooth  four-dimensional manifold which is Hausdorff, connected, and second-countable, and $g$ is a smooth Lorentzian metric on $M$ such that $(M,g)$ is time-oriented.

Our definition of timelike, null, and causal curves will follow \cite{ON}. Let $S \subset M$. The \emph{timelike future} of $S$, denoted by $I^+(S)$, is the set of points $p \in M$ such that there is a future directed timelike curve $\g \colon [a,b] \to M$ such that $\g(a) \in S$ and $\g(b) = p$. The \emph{causal future} of $S$, denoted by $J^+(S)$, is the set of points $p \in M$ such that is a future directed causal curve $\g \colon [a,b] \to M$ such that $\g(a) \in S$ and $\g(b) = p$. From Corollary 14.5 in \cite{ON}, we have

\medskip

\begin{prop}\label{null geo prop}
If $q \in J^+(p) \setminus I^+(p)$, then there is a future directed null geodesic from $p$ to $q$ without conjugate points.
\end{prop}

\medskip

A set $S \subset M$ is \emph{achronal} provided $I^+(S) \cap S = \emptyset$. The \emph{edge} of a closed and achronal set $S$, denoted by $\text{edge}(S)$, is the set of points $p \in S$ such that every neighborhood $U$ of $p$ contains points $x \in I^-(p)$ and $y \in I^+(p)$ and a timelike curve $\g \subset U$ from $x$ to $y$ such that $\g \cap S = \emptyset$. 

Let $S \subset M$ be a closed and achronal set. The \emph{future domain of dependence} of $S$, denoted by $D^+(S)$, is the set of points $p \in M$ such that every past inextendible timelike curve from $p$ intersects $S$. Note that $D^+(S)$ is a closed set \cite[Lemma 14.51]{ON}. The \emph{future Cauchy horizon} of $S$, denoted by $H^+(S)$, is the set of points $p \in D^+(S)$ such that $I^+(p) \cap D^+(S) = \emptyset$. Alternatively, we have $H^+(S) = D^+S \setminus I^-\big(D^+S)$.\footnote{To avoid clutter of parentheses, we will often abbreviate $D^+(S)$ by $D^+S$. Likewise with $I^+$ and $H^+$. }

\medskip
\medskip

\subsection{Proof of the compactness lemma}

In this section we prove the compactness lemma from \cite{BG}. Our set-up is slightly different than that of \cite{BG}, but the proofs are nearly identical. We include the proofs for the sake of completeness. 

  Recall that if $N$ is a three-dimensional connected Riemannian manifold, we say that a smooth embedded surface $S$ \emph{separates} $N$ provided $N \setminus S$ is disconnected and  $S$ is two-sided (i.e. it admits a smooth global normal vector field). 

\newpage

\noindent Consider a spacetime $(M,g)$ with the following three properties. 
\medskip

\begin{itemize}

\item[(1)] There is a smooth spacelike Cauchy hypersurface $V$ for $M$.

\item[(2)] There exists a smooth embedded surface $\S \subset V$ which separates $V$. Let $B'$ and $E'$ form a separation for $V \setminus \S$. Set $B = B'\sqcup \Sigma$ and $E = E' \sqcup \Sigma$. Hence
\[
V \,=\, B \cup E \:\:\:\: \text{ and } \:\:\:\: \S \,=\, B \cap E.
\]
We assume that $\Sigma$ is a closed set so that $B$ and $E$ are closed sets.

\item[(3)] There exists a smooth embedded 2-sphere $\Gamma \subset E \setminus \Sigma$ which separates $E \setminus \Sigma$. Let $E_1'$ and $E_2'$ form a separation for $E \setminus (\Sigma \sqcup \Gamma)$. Set $E_1 = E_1' \sqcup \Gamma \sqcup \Sigma$ and $E_2 = E_2' \sqcup \Gamma$.  Hence
\[
E \,=\, E_1 \cup E_2 \:\:\:\: \text{ and } \:\:\:\: E \,=\, E_1 \cap E_2.
\]
We assume $E_1$ is connected. We call $E_1$ and $E_2$ the \emph{inward} and \emph{outward} directions of $\Gamma$.

\end{itemize}

\medskip

\begin{Def}\label{compactness lemma late time def}
\end{Def}

\begin{itemize}

\item[$\bullet$\:\:] A spacetime $(M,g)$ \emph{has settled down at late time}  if it satisfies properties (1) -  (3) above and such that each inward pointing future inextendible null normal geodesic starting on $\Gamma$ is either future complete or crosses $H^+(E)$ (i.e.\@ intersects it transversely).

\item[$\bullet$\:\:]  Suppose $(M,g)$ has settled down at late time. We say $\Gamma$ is \emph{inner trapped} if $\theta = \div_\Gamma k < 0$; here $k = u + \nu$ where $u$ is the future directed unit normal on $V$ and $\nu$ is the inward unit normal on $\Gamma$ pointing into $E_1$. Note that $k$ is a smooth future directed inward pointing null normal vector field along $\Gamma$. 

\end{itemize}

\medskip

\noindent \emph{Remarks.} 

\begin{itemize}

\item[-] Figure \ref{set-up fig} is a good picture for our set-up provided one removes the subscript $i$.

\item[-] In the proof of Theorem \ref{main} we will apply the compactness lemma separately to each $i$. 

\item[-] We make no assumptions on the connectedness nor compactness of $\S$. In general $\S$ may consist of several (possibly infinite) connected components. 

\item[-] The proof of our theorem carries over if we assume $\Sigma$ is just $C^1$ and $\Gamma$ is just $C^2$.  We could lower the regularity of $\S$ to $C^0$ provided we also make the ``good cut" assumption made in \cite{BG}.

\item[-] $\pd B \,=\, \text{edge}(B) \,=\, \Sigma \,=\, \text{edge}(E) \,=\, \pd E.$

\item[-] $\pd E_1 \,=\, \text{edge}(E_1) \,=\, \Sigma \sqcup \Gamma$.

\end{itemize}

\medskip

\medskip

\noindent The goal of this section is to prove the following.

\medskip

\begin{lem}[Compactness lemma]\label{compactness lemma}
Suppose $(M,g)$ has settled down at late time and $\Gamma$ is inner trapped. If $(M,g)$ satisfies the null energy condition, then $E_1$ is compact.
\end{lem}

\medskip

\noindent Lemma \ref{compactness lemma} implies that $\Sigma$ can only have a finite number of components. Its proof is established through a series of propositions.

\medskip

\begin{prop}
\[
H^+(E) \,\subset\, J^+(\Sigma) \setminus I^+(\Sigma).
\]
\end{prop}

\proof
We will establish the following inclusions.

\[
H^+(E) \,\subset\, \pd I^+(B) \setminus \text{int}_V (B) \, \subset\, J^+(\Sigma) \setminus I^+(\Sigma).
\]

The left inclusion: Let $p \in H^+(E)$. Then $p \notin \text{int}_V (B)$ otherwise achronality of $V$ would be violated. Since $\Sigma \subset \pd I^+(B)$, we may assume $p \notin \Sigma$. Let $U$ be any open set about $p$. Let $x \in I^+(p) \cap U$. Then there is a past inextendible timelike curve from $x$ which does not intersect $E$. Since $V$ is a Cauchy surface, it must intersect $B$. Therefore $U$ contains a point in $I^+(B)$. Let $y \in I^-(p) \cap U$. Seeking a contradiction, suppose there is a timelike curve $\g$ from $B$ to $y$. Then $\g$ must start on $\Sigma$ otherwise we would have $p \notin D^+(E)$. Therefore $p \in I^+(\Sigma)$. But $\Sigma = \text{edge}(E)$, so we can find a past inextendible timelike curve from $p$ which doesn't meet $E$. This contradicts $p \in D^+(E)$. Therefore $y \notin I^+(B)$. Thus $p \in \pd I^+(B)$. 

The right inclusion: Let $p \in \pd I^+(B) \setminus \text{int}_V (B)$. If $p \in \Sigma$, then we are done. So assume $p \notin \Sigma$. Therefore $p \notin B$. Since $(M,g)$ is globally hyperbolic and hence causally simple, we have $\pd I^+(B) = J^+(B) \setminus I^+(B)$. Thus there exists a point $q \neq p$ with $q \in B$ and a causal curve $\g$ from $q$ to $p$. If $q \in \text{int}_V (B)$, then we can find a timelike curve $\l$ from a point $q' \in \text{int}_V(B)$ to a point on $\g$. Hence the push up lemma implies $p \in I^+(q')$, but this contradicts $p \notin I^+(B)$. Thus $q \in \Sigma$, and so $p \in J^+(\Sigma) \setminus I^+(\Sigma)$. 
\qed

\medskip

\begin{cor}\label{ruled geo cor}
Each point $p\in H^+(E)$ lies on a null geodesic $\g \subset H^+(E)$ starting from a point in $\Sigma$. 
\end{cor}

\proof
Apply Proposition \ref{null geo prop}. 
\qed

\medskip
\medskip

\begin{Def}\label{W def} \emph{We define $W$ as the closed and achronal set given by}
\[
W \,=\, (\pd I^+ E_2 \setminus \emph{int}_V E_2) \cap D^+(E).
\]
\end{Def}

\medskip

\begin{prop}\label{W compact}
Suppose $(M,g)$ has settled down at late time and $\Gamma$ is inner trapped. If $(M,g)$ satisfies the null energy condition, then $W$ is compact and meets $H^+(E)$.
\end{prop}

\proof
Put $H = H^+(E)$.  For each $x \in \Gamma$, define $\mu_x(t) = \exp_x \big(tk(x)\big)$. Recall $k$ is a smooth future directed inward pointing null normal vector field along $\Gamma$. Let $\Phi \colon \text{dom}(\Phi) \subset \Gamma \times [0,\infty) \to M$ be given by $\Phi(x,t) = \mu_x(t)$ where $\text{dom}(\Phi)$ is the maximal open set in $\Gamma \times [0,\infty)$ where $\exp$ is defined.  Define $\Gamma_H = \{x \in \Gamma \mid \mu_x \cap H \neq \emptyset\}$. A priori $\Gamma_H$ may be empty; it's only at the end of this proof can we conclude that $\Gamma_H$ is nonempty when we prove that $W$ meets $H^+(E)$. 

 Define $s \colon \Gamma_H \to \R$ by $s(x) = $ the parameter value on $\mu_x$ where $\mu_x$ meets $H$. We show $s$ is well-defined. If $x \in \Gamma_H$, then there exists a $t > 0$ such that $\mu_x(t) \in H$. We know $\mu_x(t) \notin \Sigma$ otherwise we could violate the achronality of $V$ via the push up lemma. Since $H$ is ruled by null geodesics by Corollary \ref{ruled geo cor}, we can find a point $y \in H \setminus \S$ and a null geodesic on $H$ from $y$ to $\mu_x(t)$. Since $\mu_x$ must cross $H$ (by our late time assumption), there exists an $\e > 0$ such that $\mu_x(t + \e') \notin H$ for all $0 < \e' < \e$. By choosing $\e'$ small enough and $y$ sufficiently close to $\mu_x(t)$, we can find a timelike curve from $y$ to $\mu_x(t + \e')$. Therefore, if there exists a time $t_0 > t$ such that $\mu_x(t_0) \in H$, then we can find a future directed timelike curve from $y$ to $\mu_x(t_0)$ by the push up lemma, but this contradicts the achronality of $H$. Thus $s$ is well-defined.

 Now we show $s$ is continuous. Let $x_n$ be a sequence of points in $\Gamma_H$ which converges to $x \in \Gamma_H$. For each $x_n$, there exists a $t_n$ such that $\mu_{x_n}(t_n) \in H$ (i.e. $s(x_n) = t_n$). Since $x \in \Gamma_H$, there exists a $t$ such that $\mu_x(t) \in H$. Fix $\e > 0$. Then $\mu_x(t + \e) \notin D^+(E)$ and since $D^+(E)$ is closed, there is an open set $U$ about $\mu_x(t + \e)$ such that $U \cap D^+(E) = \emptyset$. Let $V = \Phi^{-1}(U)$. Then $\pi(V)$ is open where $\pi \colon \Gamma \times [0,\infty) \to \Gamma$ is the projection map. Hence, there is an $N$ such that $x_n \in \pi(V)$ for all $n > N$.  Consequently $\mu_{x_n}(t + \e) \in U$ for all $n > N$. Since $U \cap D^+(E) = \emptyset$, it follows that $t_n \leq t + \e$. Similarly, $\mu_x(t - \e) \in \text{int} D^+(E)$ for all sufficiently small $\e > 0$. Therefore a similar argument shows there exists an $N'$ such that $t_n \leq t - \e$ for all $n > N'$. Thus $t_n \to t$, and so $s$ is continuous. 
 
 For each $x \in \Gamma$, let $r(x) = 2/|\theta(x)|$ where $\theta(x) = \div_\Gamma k(x)$. Since $\Gamma$ is compact, we define $s_0 = \max_{x \in \Gamma} \{r(x)\}$. Note that if $\mu_x$ extends to $s_0$, then $\mu_x|_{[0,s_0]}$ contains a null focal point of $\Gamma$ \cite[Prop. 10.43]{ON}. Let $\Gamma_0 = s^{-1}\big([0,s_0]\big)$.  We define the function $\hat{s}\colon \Gamma \to \R$ by $\hat{s}(x) = s(x)$ if $x \in \Gamma_0$ and $s(x) = s_0$ if $x \notin \Gamma_0$. Continuity of $s$ implies $\pd_{\Gamma} \Gamma_0 \subset s^{-1}(s_0)$ and this implies $\hat{s}$ is continuous. Let $A = \big\{(x,t) \in \Gamma \times [0, \infty) \mid t \in \big[0, \hat{s}(x)\big]\big\}$. Continuity of $\hat{s}$ and compactness of $\Gamma$ implies $A$ is compact. Note that $A \subset \text{dom}(\Phi)$ by our late time assumption and hence $\Phi(A)$ makes sense.
 
Now we show $W \subset \Phi(A)$ from which compactness of $W$ follows since $W$ is closed. Fix $x \in W$. Then $x \in \pd I^+ (E_2)$ implies $x \in J^+ (E_2) \setminus I^+ (E_2)$. Therefore the push up lemma implies $x \in J^+(\Gamma)$. Hence $x \in J^+(\Gamma) \setminus I^+(\Gamma)$, and so there is a null geodesic $\mu$ from a point $x_0 \in \Gamma$ to $x = \mu(t)$ for some $t \geq 0$. If $t > \hat{s}(x_0)$, then either $\mu$ leaves $D^+(E)$ by crossing $H$ or leaves $\pd I^+(E_2)$ by encountering a null focal point \cite[Prop. 10.48]{ON}. In either case we would have $\mu(t) \notin W$ which is a contradiction. Therefore $t \leq \hat{s}(x_0)$ which implies $(x_0, t) \in A$ which implies $x = \Phi(x_0, t) \in \Phi(A)$. Hence $W \subset \Phi(A)$.
 
Lastly we show $W$ meets $H$. If $W$ does meet $H$, then it meets $H$ transversely by our late time assumption. This implies that $\text{edge}(W) = \Gamma \sqcup (H \cap W)$. Seeking a contradiction, suppose $W$ does not meet $H$. Then $\text{edge}(W) = \Gamma$. Let $X$ be a past directed smooth timelike vector field on $M$. For each $p \in W$, the integral curve of $X$ passing through $p$ meets $E$ in a unique point $\tau(p)$. Note $\tau (p) \notin \text{int}_VE_2$ otherwise there would be a timelike curve from $E_2$ to $W$. This defines a flow map $\tau \colon W \to E_1$ which is injective since integral curves do not intersect. Set $W' = W \setminus \Gamma$. Since $W'$ is a $C^0$ hypersurface \cite[Prop. 14.25]{ON}, we have $\tau|_{W'}$ is an open map by Brower's invariance of domain theorem. Since $\tau$ is just the identity on $\Gamma$, we have $\tau$ is an open map too. Hence $\tau(W)$ is open. Also, $\tau(W)$ is closed by compactness of $W$. Since $E_1$ is connected, we have $\tau(W) = E_1$. Therefore there is a timelike curve from a point on $\Sigma$ to a point $q \in W$. But since $\Sigma = \text{edge}(E)$, we can find a past inextendible timelike curve from $q$ which does not meet $E$. This is a contradiction.
\qed

\medskip
\medskip

\noindent\emph{Remark.} Set $\Gamma = \Gamma_1$ in Figure \ref{SdS fig}. In this case $W$ meets the event horizon as in the conclusion of Proposition \ref{W compact}. Now imagine that $\Gamma$ lies past the cosmological horizon (which is \emph{the opposite} of what is shown in Figure \ref{SdS fig}), then the future light cone of $\Gamma$ does not intersect the event horizon. Proposition \ref{W compact} implies that $\Gamma$ cannot be inner trapped in this case. 

\medskip
\medskip

\begin{prop}\label{horizon E1 prop}
\[H^+(E_1) \,\subset\, H^+(E) \cup W.\]
\end{prop}

\proof
Again set $H = H^+(E)$. We have
\[
H^+(E_1) \,\subset\, D^+(E_1) \,\subset\, D^+(E) \,=\, \pd (D^+E )\cup \text{int}(D^+E) \,=\, H \cup E \cup \text{int}(D^+E).
\]
The last equality follows from \cite[Lemma 10.52]{ON}. Fix $p \in H^+(E_1)$. If $p \in H$, then we are done. If $p \in E$, then $p \in E_1$ by achronality. Hence $p \in \text{edge}(E_1) = \Sigma \cup \Gamma \subset H \cup W$. Lastly, suppose $p \in \text{int}(D^+E)$. Then $p \notin E_2$ since $E_2 \subset \pd D^+(E)$. Let $U \subset \text{int}(D^+E)$ be any open set around $p$. We have $p \notin I^+(E_2)$  otherwise we could find a timelike curve from $\text{int}_VE_2$ to $p$ which contradicts $p \in H^+(E_1)$. Thus $U$ contains a point not in $I^+(E_2)$. Let $q \in I^+(p) \cap U$. Then $q \in D^+(E) \setminus D^+(E_1)$ which implies $q \in I^+(E_2)$. Thus $U$ contains a point in $I^+(E_2)$. Therefore $p \in \pd I^+(E_2)$. Hence $p \in W$.
\qed

\medskip
\medskip

\begin{Def}
\emph{
Let $H^+_\Gamma(E_1)$ be the connected component of $H^+(E_1)$ containing $\Gamma$. 
}
\end{Def}

\medskip

\begin{prop}\label{H^+ is cpt prop}
Suppose $(M,g)$ has settled down at late time and $\Gamma$ is inner trapped. If $(M,g)$ satisfies the null energy condition, then $H_\Gamma^+(E_1)$ is compact.
\end{prop}

\proof
Again set $H = H^+(E)$. Let $H_0$ be the union of components of $H$ which meet $W$ which is nonempty by Proposition \ref{W compact}. Let $\Sigma_0 = \Sigma \cap H_0$.  By Proposition \ref{horizon E1 prop}, we have $H^+_\Gamma(E_1) \subset W \cup H$, but since $W \cup H_0$ is connected, we have $H^+_\Gamma(E_1) \subset W \cup H_0$. We will show that there is in fact a compact subset $H_0' \subset H_0$ such that $H^+_\Gamma(E_1) \subset W \cup H_0'$. Then compactness will follow since $H^+_\Gamma(E_1)$ is closed. 

By Corollary \ref{ruled geo cor}, $H_0$ is ruled by null geodesics emanating from $\Sigma_0$. Claim: Each such null geodesic meets $W$ exactly once. To prove the claim, let $n$ be a smooth outward pointing unit normal vector field along $\S_0$ (outward pointing is with respect to $E_1$). Let $\ell = u + n$ where $u$ is the future directed unit normal to the Cauchy surface $V$. Define $\eta_x(t) = \exp_x t\ell(x)$. Let $\Phi \colon \text{dom}(\Phi) \subset \Sigma_0 \times [0,\infty) \to M$ be given by $\Phi(x,t) = \eta_x(t)$ where $\text{dom}(\Phi)$ is the maximal open set in $\Sigma_0 \times [0,\infty)$ where $\exp$ is defined. Let $\pi \colon \Sigma_0 \times [0, \infty) \to \Sigma_0$ denote the projection map. Set $A = \Phi^{-1}(W \cap H_0)$. Then $A$ is compact since $W$ is compact and $H_0$ is closed. Therefore $\pi(A)$ is compact and hence closed. Now we prove $\pi(A)$ is open. Fix $x \in \pi(A)$. Then there is a $t > 0$ such that $\eta_x(t) \in W$. Since $W$ meets $H_0$ transversely, there exists an $\e > 0$ such that $\eta_x(t + \e') \in I^+(E_2)$ for all $0 < \e' \leq \e$. Let $U \subset I^+(E_2)$ be an open set about $\eta_x(t + \e)$. Let $V = \Phi^{-1}(U)$. Then $\pi(V) \subset \Sigma_0$ is open about $x$. For every $y \in \pi(V)$, there exists a $t_y$ such that $\eta_y(t_y) \in U$. Since $\eta_y(0) \notin I^+(E_2)$ and $\eta_y(t_y) \in I^+(E_2)$, there exists a $t_y' \in (0, t_y)$ such that $\eta_y(t_y') \in \pd I^+(E_2)$. Since $\eta_y \subset H \subset D^+(E)$, we have $\eta_y(t_y') \in W$. 
Thus $\pi(V) \subset \pi(A)$ and so $\pi(A)$ is open. Since $\pi(A)$ is both open and closed, we have $\pi(A)$ maps onto each connected component of $\Sigma_0$. Thus $\pi(A) = \Sigma_0$. Therefore we have proved that each null geodesic on $H_0$ meets $W$. Now suppose $\eta$ is a null geodesic on $H_0$ which meets $W$ at $\eta(t_1)$ and $\eta(t_2)$ with $t_2 > t_1$. Since $W$ meets $H$ transversally, we can find a broken null geodesic from a point on $\Gamma$ to $\eta(t_2)$ where the break point occurs at $\eta(t_1)$. By the push up lemma, we can find a timelike curve from $\Gamma$ to $\eta(t_2)$ which implies $\eta(t_2) \in I^+(E_2)$. This is a contradiction. Therefore the claim is proved. 

Define $s \colon \Sigma_0 \to \R$ by $s(x) =$ the parameter value on $\eta_x$ where $\eta_x$ meets $W$. By the above claim, we have $s$ is well-defined. Moreover $s$ is continuous by an analogous argument as used in the proof of Proposition \ref{W compact}. Define $C = \big\{(x,t) \in \Sigma_0 \times [0,\infty) \mid t \in [0, s(x)]\big\}$. Then $C$ is compact which follows from continuity of $s$ and compactness of $\Sigma_0$; note that $\Sigma_0$ is compact follows from $\Sigma_0 = \pi(A)$. Then $H_0' = \Phi(C)$ is compact. Now observe that $H_0 \setminus H_0' \subset I^+(E_2)$. Hence $H^+(E_1) \cap (H_0 \setminus H_0') = \emptyset$. Thus $H^+_\Gamma(E_1) \subset W \cup H_0'$.
\qed

\medskip
\medskip

\noindent\underline{\emph{Proof of Lemma \emph{\ref{compactness lemma}} (Compactness Lemma)}}.

\medskip
Using the notation from the proof of Proposition \ref{H^+ is cpt prop}, note that $\text{edge}\big(H^+_\Gamma(E_1)\big) = \Sigma_0 \sqcup \Gamma$. Let $H^+_\Gamma(E_1)'$ denote $H^+_\Gamma(E_1) \setminus (\Sigma_0 \sqcup \Gamma)$ which is a $C^0$ hypersurface \cite[Prop. 14.25]{ON}.

Let $X$ be a past directed smooth timelike vector field on $M$. For each $p \in H^+_\Gamma(E_1)$, the integral curve of $X$ passing through $p$ meets $E_1$ in a unique point $\tau(p)$. This defines a flow map $\tau \colon H^+_\Gamma(E_1) \to E_1$.  Since integral curves do not intersect, $\tau$ is injective.  Therefore $\tau$ restricted to $H^+_\Gamma(E_1)'$ is an open map by Brower's invariance of domain theorem. Since $\tau$ is just the identity on $\S_0 \sqcup \Gamma$, we have $\tau$ is an open map too. Hence the image of $\tau$ is open in $E_1$. Also the image of $\tau$ is closed in $E_1$ by compactness of $H^+_\Gamma(E_1)$. Since $E_1$ is connected, the image of $\tau$ is all of $E_1$. Thus $E_1$ is compact. 
\qed

\medskip
\medskip

\noindent\emph{Remark.} From the compactness lemma, it follows that $\Sigma_0 = \Sigma$ and $H_0 = H$. To see this, note that the proof of the compactness lemma shows $\text{edge}(H^+_\Gamma(E_1)\big) = \text{edge}(E_1)$. Hence $\Sigma_0 \sqcup \Gamma = \Sigma \sqcup \Gamma$. Therefore $\Sigma_0 = \Sigma$ and $H_0 = H$.

\medskip
\medskip

\subsection{Proof of Theorem \ref{main}}\label{proof section}

To prove Theorem \ref{main} we will apply the compactness lemma in a suitably constructed covering spacetime. 
%
The covering argument uses similar ideas to that in \cite{BG}, but in \cite{BG} there is a single $\Gamma$ (no subscript $i$) and the subset $E_2$ is assumed to be noncompact; we cover this case in Theorem \ref{main}(b)(ii).  Although this makes the contradiction with the compactness lemma immediate, it blurs some interesting cases that we will cover here. By splitting Theorem \ref{main} into the subcases $N=1$ and $N\geq 2$, we can remove the assumption in \cite{BG} that $E_2$ is noncompact for $N\geq 2$. As for $N=1$, we allow for $E_2$ to be compact and obtain additional conclusions in that setting. From a physical perspective, the case $N = 1$ could correspond to the situation when one envelops all the black holes in the universe within a (very large) sphere $\Gamma$. From a more mathematical perspective, the case $N = 1$ corresponds to the case when there is a single black hole in an otherwise empty universe (e.g. a single black hole in an asymptotically flat spacetime).

%

\medskip
\medskip

\noindent\underline{\emph{Proof of Theorem \emph{\ref{main}(a)}}}.

\medskip

 Fix $i \in \mathcal{I}$. We first show $E_{1,i}$ is compact. This follows from an application of the compactness lemma (Lemma \ref{compactness lemma}) with 
 \[
V,\, \S_i,\, \Gamma_i,\, B_i,\, E_i,\, E_{1,i}, E_{2,i}\, \text{ playing the roles of }\, V,\, \S,\, \Gamma,\, B,\, E,\, E_1,\, E_2,
\]
respectively. If $E_{1,i}$ is simply connected, then \cite[Lemma 4.9]{Hempel} implies that $\S_i$ is a finite disjoint union of $S^2$s. Then $\pi_1(V) = \pi_1(C)$ by the Seifert-Van Kampen theorem. Thus it suffices to show that $E_{1,i}$ is simply connected for each $i$; we show this next.

Set $F_i = B_i \cup E_{1,i}$ so that $V = F_i \cup E_{2,i}$. Let $\wt{F}_i$ denote its universal cover with covering map $p \colon \wt{F}_i \to F_i$. Set $\wt{B}_i = p^{-1}(B_i)$ and likewise with $\wt{E}_{1,i}$, $\wt{\Sigma}_i$, and $\wt{\Gamma}_i$. Since $E_{1,i}$ is a deformation retract of $F_i$, it follows that $\wt{E}_{1,i}$ is the universal cover of $E_{1,i}$ with covering map given by the restriction of $p$. Since $\Gamma_i$ is simply connected, $\wt{\Gamma}_i$ is the disjoint union of $n$ copies of $\Gamma_i$ where $n$ is the number of sheets (possibly infinite) in the covering space $\wt{E}_{1,i}$. We write this as $\wt{\Gamma}_i = \bigsqcup_{\a \in A} \Gamma_i^\a$ where $A$ is an indexed set with cardinality $n =|A|$. Take $n$ copies of $E_{2,i}$, call them $E_{2,i}^\a$ for $\a \in A$, and glue each copy onto $\wt{F}_i$ by identifying the boundary of $E_{2,i}^\a$ with the boundary component $\Gamma^\a_i$ in the same way they're attached in the base space $V$. Let $\wt{V}$ denote the resulting Riemannian manifold. Then $\wt{V}$ is a covering of $V$, and abusing notation, we still call this covering map $p \colon \wt{V} \to V$. Set $\wt{E}_{2,i} = p^{-1}(E_{2,i}) = \bigsqcup_{\a \in A} E_{2,i}^\a$. Using covering space theory (as outlined in \cite{BG}) or the Bernal-Sanchez splitting result (as outlined in \cite{GL}), there is a covering spacetime $\wt{M}$ of $M$ with a localy isometry covering map $P \colon \wt{M} \to M$ such that $\wt{V}$ is a Cauchy surface for $\wt{M}$ and $P|_{\wt{V}} = p$.  

Now we wish to apply the compactness lemma  to the covering spacetime. However, the sets $\wt{E}_{1,i}$ and $\wt{E}_{2,i}$ do \emph{not} play the roles of $E_1$ and $E_2$. These will be played by $D_1$ and $D_2$ which we define now. Let $\Gamma_i^{\bar{\a}}$ denote a single component of $\wt{\Gamma}_i = \bigsqcup_{\a \in A} \Gamma_i^\a$. Then $\Gamma_i^{\bar{\a}}$ separates $\wt{E}_i \setminus \wt{\S}_i$ where $\wt{E}_i = p^{-1}(E_i)$. Let $D_1'$ and $D_2'$ form a separation for $\wt{E}_i \setminus (\wt{\S} \sqcup\Gamma_i^{\bar{\a}})$. Set $D_1 = D_1' \sqcup \Gamma_i^{\bar{\a}} \sqcup \wt{\Sigma}_i$ and $D_2 = D_2' \sqcup \Gamma_i^{\bar{\a}}$. Note that $D_2$ is topologically $E_{2,i}$. $D_1$ is connected since $E_{1,i}$ is connected which implies $\wt{E}_{1,i}$ is connected which implies $D_1$ is connected. We will apply the compactness lemma with 
\[
\wt{V},\, \wt{\S}_i,\, \Gamma_i^{\bar{\a}},\, \wt{B}_i,\, \wt{E}_i,\, D_1, D_2\, \text{ playing the roles of }\, V,\, \S,\, \Gamma,\, B,\, E,\, E_1,\, E_2,
\]
respectively.  $\Gamma_i^{\bar{\a}}$ is inner trapped since the covering map $P \colon \wt{M} \to M$ is a local isometry. Now we show that the late time assumption in Definition \ref{compactness lemma late time def} holds: Let $\gamma$ be an inward pointing future inextendible null normal geodesic starting on $\Gamma^{\bar{\a}}_i$ which is not future complete. Then $P \circ \gamma$ is not future complete. Therefore $P \circ \gamma$ crosses $H^+(E_i)$. Therefore $\gamma$ crosses $H^+(\wt{E}_i)$.

Thus we can apply the compactness lemma. Seeking a contradiction, assume $E_{1,i}$ is not simply connected. Then $n \geq 2$. Therefore $D_1$ contains at least one copy of $E_{2,i}$ attached to some $\Gamma_i^\a \neq \Gamma_i^{\bar{\a}}$. Since we're assuming $N \geq 2$, there is an index $j \in \mathcal{I}$ such that $j \neq i$, and so $E_{2,i}$ contains a copy of $B_j$. Therefore $D_1$ contains a copy of $B_j$. Since $B_j$ is topologically $\S_j \times [0, \e)$, we have $D_1$ is not compact which contradicts the compactness lemma.
\qed

\medskip
\medskip

\noindent\underline{\emph{Proof of Theorem \emph{\ref{main}(b)}}}.
\begin{itemize}
\item[(i)] Consider the cover $\wt{V}$ of $V$ constructed in the proof of part (a), but now we don't need the subscript $i$. This construction yields the sets 
\[
\wt{V},\, \wt{\S},\, \Gamma^{\bar{\a}},\, \wt{B},\, \wt{E},\, D_1, D_2\, \text{ which play the roles of }\,  V,\, \S,\, \Gamma,\, B,\, E,\, E_1,\, E_2,
\]
in the compactness lemma. If $\pi_1(E_1)$ was infinite, then $D_1$ would be noncompact since it contains $\wt{E}_1$. This contradicts the compactness lemma. That $\S$ is a finite disjoint union of $S^2$s follows in the same way as in the proof of part (a). 

\item[(ii)] Again consider the cover $\wt{V}$ of $V$ constructed in the proof of part (a). Let $n$ denote the cardinality of $\pi_1(E_1)$. If $\pi_1(E_1)$ is not trivial, then $n \geq 2$. Therefore $D_1$ contains at least one copy of $E_2$ attached to some $\Gamma^\a \neq \Gamma^{\bar{\a}}$. If $E_2$ is noncompact, then $D_1$ is noncompact which contradicts the compactness lemma.

\item[(iii)] Assume $\pi_1(E_2)$ is nontrivial. Let $F = B \cup E_1$. Then $\Gamma = \pd F = \pd E_2$.  Let $\wt{F}$ and $\wt{E}_2$ denote the universal covers of $F$ and $E_2$ with covering maps $p_1 \colon \wt{F} \to F$ and $p_2 \colon \wt{E}_2 \to E_2$.  Let $n$ and $m$ denote the number of sheets in $\wt{F}$ and $\wt{E}_2$, respectively. Since $\Gamma$ is simply connected, we have $p_1^{-1}(\pd F) = \bigsqcup_{\a \in A} \Gamma^\a$ and $p_2^{-1}(\pd E_2) = \bigsqcup_{\beta \in B} \Gamma^\beta$ where $A$ and $B$ are indexed sets with cardinalities $n$ and $m$, respectively. For each $\a \in A$, let $\wt{E}_2^\a$ denote a copy of $\wt{E}_2$. Let $\bigsqcup_{\a, \b} \Gamma^\b_\a$ denote the boundary of $\bigsqcup_{\a} \wt{E}_2^\a$. Let $\Gamma^{\ov{\b}}$ denote a single component of $\bigsqcup_{\beta}\Gamma^\b$. Let $\Gamma^{\ov{\b}}_\a$ denote the same component within $\bigsqcup_{\a,\b}\Gamma^\b_\a$. Glue $\wt{F}$ and $\bigsqcup_{\a} \wt{E}_2^\a$ together by attaching $\Gamma^{\ov{\b}}_\a$ to $\Gamma^\a$ in the same way they're attached in the base space $V$; we do this gluing for each $\a \in A$. The resulting space is a manifold with boundary. Now for each $\a$ and for each $\b \neq \ov{\b}$, we glue $F$ to $\Gamma^\b_\a$ along $\Gamma = \pd F$. The resulting space is a Riemannian manifold $\wt{V}$ (without boundary). Let $\Gamma^{\bar{\a}}$ denote a single component of $\bigsqcup_{\a} \Gamma^\a$. Let $p \colon \wt{V} \to V$ denote the covering map (hence $p_1$ and $p_2$ are restrictions of $p$ to $\wt{F}$ and $\wt{E}_2$, respectively). Set $\wt{\S} = p^{-1}(\S)$ and likewise with $\wt{B}$ and $\wt{E}$. Let $D_2$ denote $\wt{E}_2^{\bar{\a}}$ (which is attached to $\Gamma^{\bar{\a}}$). Let $D_1$ denote $(\wt{V} \setminus D_2) \sqcup \Gamma^{\bar{\a}}$. Then we apply the compactness lemma with
\[
\wt{V},\, \wt{\S},\, \Gamma^{\bar{\a}},\, \wt{B},\, \wt{E},\, D_1, D_2\, \text{ playing the roles of }\, V,\, \S,\, \Gamma,\, B,\, E,\, E_1,\, E_2,
\]
respectively. Seeking a contradiction, assume $\pi_1(E_1)$ is nontrivial. Then $n \geq 2$. Therefore there is some $\a$ such that $\Gamma^\a \neq \Gamma^{\bar{\a}}$. Therefore $D_1$ contains $\wt{E}_2^\a$. Since $\pi_1(E_2)$ is nontrivial, $m \geq 2$. Therefore $\pd \wt{E}_2^\a$ contains a component $\Gamma_\a^{\b} \neq \Gamma_{\a}^{\ov{\b}}$. Since $F$ is glued along $\Gamma_\a^{\b}$, it follows that $D_1$ contains a copy of $B$. Since $B$ is topologically $\S \times [0, \e)$, it follows that $D_1$ is noncompact. But this contradicts the compactness lemma. 

\item[(iv)] Assume $\pi_1(E_1)$ is nontrivial. By parts (ii) and (iii), we have $E_2$ is simply connected and compact. Smoothly attach a 3-disc to $E_2$ along $\Gamma$. The resulting space, call it $S$, is simply connected by the Seifert-Van Kampen theorem. Thus $S$ is a closed and simply connected 3-manifold. Therefore $S$ is topologically $S^3$ by the positive resolution of the Poincar{\'e} conjecture. $E_2$ is then the complement of a 3-disc with its boundary removed in $S^3$. Hence $E_2$ is a 3-disc by Alexander's theorem \cite[Thm. 1.1]{Ha}.  \qed
\end{itemize}

\newpage

\bibliographystyle{amsplain}
\bibliography{ds top cen}

\begin{thebibliography}{99}
\bibitem{BG} 
S. Browdy and G. Galloway, \emph{Topological censorship and the topology of black holes}, Journal of Mathematical Physics \textbf{36}, (1995).

\bibitem{ChruBook}
P. Chru{\'s}ciel, \emph{Geometry of Black Holes},  Oxford University Press,  Oxford, (2020). 

\bibitem{CG19} 
P. Chr{\'u}sciel and G. Galloway, \emph{Roads to topological censorship}, arXiv:1906.02151 (2019).

\bibitem{CGS}
P. Chru{\'u}sciel, G. Galloway, and D. Solis, \emph{Topological censorship for Kaluza-Klein space-times}, Annales Henri Poincare \textbf{10}, (2009).

\bibitem{CMa}
P. Chr{\'u}sciel and R. Mazzeo, \emph{On ``many black hole" vacuum spacetimes}, Classical and Quantum Gravity \textbf{20} (2003).


\bibitem{FSW93} 
J. Friedman, K. Schleich, and D. Witt, \emph{Topological censorship}, Physical Review Letters \textbf{71} (1993), erratum \textbf{75} (1995).

\bibitem{G95} 
G. Galloway, \emph{On the topology of the domain of outer communication}, Classical and Quantum Gravity \textbf{12} (1995).

\bibitem{G96} 
G. Galloway, \emph{A ``finite infinity" version of the FSW topological censorship}, Classical and Quantum Gravity \textbf{13} (1996).

\bibitem{GGL}
G. Galloway, G. Graf, E. Ling, \emph{A conformal infinity approach to asymptotically $\emph{\text{AdS}}_2\times S^{n-1}$ spacetimes}, Annales Henri Poincar{\'e} {\bf 21} (2020). 

\bibitem{GL} 
G. Galloway and E. Ling, \emph{Topology and Singularities in Cosmological Spacetimes Obeying the Null Energy Condition}, Communications in Mathematical Physics \textbf{360} (2017).


\bibitem{GSWW99} 
G. Galloway, K. Schleich, D. Witt, and E. Woolgar, \emph{Topological censorship and higher genus black holes}, Physical Review D \textbf{60} (1999).

\bibitem{GW96} 
G. Galloway and E. Woolgar, \emph{The cosmic censor forbids naked topology}, Classical and Quantum Gravity \textbf{14} (1997).

\bibitem{HE}
S. Hawking and G. Ellis, \emph{The large-scale structure of space-time}, Cambridge University Press, London, (1973). 

\bibitem{Ha}
A. Hatcher, \emph{Notes on Basic $3$-Manifold Topology}.

\bibitem{Hempel}
J. Hempel, \emph{3-manifolds}, Princeton University press, Princeton, NJ, (1976).

\bibitem{ON} 
B. O'Neill, \emph{Semi-{R}iemannian geometry}, Pure and Applied Mathematics,
  vol. 103, Academic Press Inc. [Harcourt Brace Jovanovich Publishers], New
  York, (1983).

\bibitem{Planck}
Planck Collaboration, \emph{Planck 2018 results}, Astronomy and Astrophysics \textbf{641} (2020).

\bibitem{Wald}
R. Wald, \emph{General Relativity}, University of Chicago Press, Chicago, IL, (1984).

\end{thebibliography}

\end{document}